# In Search of Simplicity: A Self-Organizing Multi-Source Multicast Overlay

Matei RIPEANU, Ian FOSTER, Adriana IAMNITCHI, Anne ROGERS

***Abstract*--Multicast communication primitives have broad utility as building blocks for distributed applications. The challenge is to create and maintain the distributed structures that support these primitives while accounting for volatile end-nodes and variable network characteristics. Most solutions proposed to date rely on complex algorithms or global information, thus limiting the scale of deployments and acceptance outside the academic realm.**

**This article introduces a low-complexity, self-organizing solution for maintaining multicast trees, that we refer to as UMM (Unstructured Multi-source Multicast). UMM uses traditional distributed systems techniques: layering, soft-state, and passive data collection to adapt to the dynamics of the physical network and maintain data dissemination trees. The result is a simple, adaptive system with lower overheads than more complex alternatives. We have implemented UMM and evaluated it on a 100-node PlanetLab testbed and on up to 1024-node emulated ModelNet networks Extensive experimental evaluations demonstrate UMM's low overhead, efficient network usage compared to alternative solutions, and ability to quickly adapt to network changes and to recover from failures.**

## I INTRODUCTION

Collaborative applications such as conferencing, shared virtual workspaces, or multi-player games rely on multi-source multicast functionality. Recently, this functionality has been used in contexts such as peer-to-peer resource discovery [1] or resource monitoring [2]. Limited native multicast support in the Internet [3] and application-specific requirements (e.g., reliability, message stream semantics) make application overlays an attractive alternative to IP-layer multicast [4-9]. Application overlays [4] employ participating end-systems to implement functionality not provided by the underlying communication layer or to improve the characteristics of services provided by lower network layers. When providing multicast functionality, overlays serve as support for extracting dissemination trees used to send data from each source to all destinations.

Two broad classes of overlays have emerged, differentiated by the existence of global rules (or the lack of thereof) that restrict their geometry. *Unstructured overlays* allow unrestricted interaction patterns between participating nodes, while *structured overlays* impose global, regular structures. Intuitively, unstructured overlays are less expensive to create and maintain as there are no topological restrictions. Unstructured overlays have more flexibility to map on inherently heterogeneous sets of end-nodes linked by a physical network, the Internet, that displays scale-free [10] and small-world [11] characteristics. However, existing solutions for multi-source multicast based on unstructured overlays have limited scalability [12, 13] or incur large overheads [1]. On the other side, the scalability advantage offered by structured overlays [14, 15] is offset by less efficient usage of underlying network resources and high protocol complexity [16].

The structured vs. unstructured overlay debate is particularly relevant in a large-scale data distribution context because of the tension between scalability and efficient network usage. Our research brings new evidence to this debate by proposing a multi-source multicast solution based on unstructured overlays and extensively evaluating it against the best solutions proposed so far in both the structured and unstructured overlay classes.

While novel as a whole solution, the multi-source multicast infrastructure we propose builds on traditional, well understood building blocks. We have reached this design in our quest for *simplicity*. We use recently proposed heuristics [17, 18] to build and optimize an unstructured base overlay. On top of it, we use flood-and-prune techniques to select efficient, source-specific multicast distribution trees: we use the implicit information contained in the duplicate messages that result from flooding to filter out redundant overlay paths. In the following, we refer to the resulting Unstructured Multi-source Multicast solution as UMM. This solution:

- has *low complexity*: it relies on soft-state and passive data collection to extract multicast dissemination trees and adapt to participating end-host and physical network dynamics. We assert that reduced complexity is a highly desirable property of distributed, large-scale systems.
- is *self-organizing*: independent decisions made at each node based on partial, local information result in desired global system behavior.
- is *scalable* with the number of sources and independent of the number of participants.
- *recovers* quickly from significant failures in the set of participating end-hosts and *adapts* to changes in underlying network topology with minimal disruption to the service offered.

- is *independent of the topology of the base overlay* and of the mechanisms used to build it which makes it reusable in a wide array of situations.

This work makes multiple contributions. First, we combine in a novel way a number of traditional, distributed systems techniques to build a simple, efficient multi-source multicast overlay. Second, and more importantly, we show, via extensive experimental evaluations over live a live wide-area testbed (PlanetLab [19]) and large emulated networks (ModelNet [20]), that although UMM uses a low-complexity protocol and nodes maintain little state, the resulting dissemination trees are more efficient than those offered by structured overlay-based solutions or by unstructured overlay-based solutions that maintain global state. Additionally, we confirm previous results regarding the relative performance of alternative overlay solutions obtained only through low-fidelity simulations by evaluating actual implementations over large-scale emulated networks. Third, we identify optimal overlay layouts and use them to quantify the overheads introduced by UMM and alternative solutions. This offers an upper bound for potential gains achievable through better decentralized overlay construction algorithms. Fourth, we present an empirical method for choosing the parameters that drive the tradeoff between agility to adapt and stability, a tradeoff all systems that need to adapt to the evolving characteristics of their underlying platform have to decide on.

The rest of this article is organized as follows. Section II surveys the main challenges application-layer overlays face and the metrics employed to evaluate how well particular solutions meet these challenges. Section III presents the main alternative overlay topologies used to provide multi-source multicast functionality. Section IV presents the UMM design. Implementation details are presented in Section V and experimental evaluations in Section VI. Section VII surveys related work and Section VIII concludes.

## II OVERVIEW

In this section we discuss challenges and success metrics associated with building a multi-source multicast solution, state our design choices, and sketch our solution.

### *A. Challenges and Success Metrics*

This section identifies the key challenges, and for each challenge, it also identifies the metrics used to evaluate success and to compare with alternative approaches.

*i.) Efficient usage of underlying network resources.* Data dissemination trees should map well on the underlying network. Since overlays are implemented on top of IP, the efficiency of the dissemination trees used should be compared with IP-multicast. An efficient tree minimizes: *relative delay penalty* (RDP), i.e., the ratio between the message propagation delay in the overlay and in the underlying network; and *network stress*, i.e., the number of duplicate messages generated due to mapping multiple overlay tunnels over the same physical link. These metrics are defined with respect to a common IP-layer multicast base and have been traditionally used to evaluate multicast solutions.

*ii.) Scalability* is obtained by minimizing the overheads to maintain efficient distributed structures: the base overlay and data dissemination trees in this case. Two main types of overheads can be identified: overheads for estimating the conditions of the underlying network (such as latency or available bandwidth) and state maintenance overheads (such as routing table updates, connectivity maintenance, partition detection, and overlay bootstrap control messages).

*iii.) Resilience*. When not bandwidth-constrained, a multicast infrastructure should deliver each message to all destinations in spite of node failures or adaptation events. To quantify resilience, we estimate *delivery rates* under various failure scenarios.

*iv.) Adaptation*. The infrastructure should build and maintain efficient delivery trees in spite of end-host failures and changes in the topology and state of the underlying network. Moreover, the overlay should adapt quickly to changes in the properties of participating resources and end-host and routing failures. If adaptation is a process by which a change (e.g., failure) is responded to, we can measure the 'success' of adaptation by tracking the evolution of metrics that characterize overlay efficiency (network stress, RDP) or delivered data rates. Presumably, we will typically see these metrics first drop and then increase to a new stable point. The time required to cross a certain threshold of effectiveness is a measure of *agility*, as is the time required to stabilize.

*B. Design Choices*

We summarize the salient design choices of our solution:

*i.) Best effort service:* Ideally, all nodes receive all transmitted messages. However, messages may be lost due to node crashes, overlay self-organization events, or insufficient transport capacity in the underlying network. UMM aims to provide a best effort service; reliable delivery and flow and congestion control can be implemented at a higher layer, if necessary. UMM, as most other solutions, does not attempt to provide these high-level properties because their semantics are generally application specific.

*ii.) Decentralized design:* Solutions that rely heavily on centralized, stable components have higher adoption barriers due to the need to define, deploy, and operate these components. Thus, we seek a solution with minimal dependence on centralized, stable components. Moreover, we attempt to minimize the state maintained at each participating host in order to limit state maintenance overheads.

*iii.) Scale:* We target *medium scale* groups with thousands rather than millions of sources. This restriction however does not apply to the number of passive participating nodes.

## *C. Solution Overview*

UMM builds two layers on top of the physical network (Figure 1). First, the *base overlay*: the set of all logical *tunnels* that connect pairs of overlay nodes using physical network paths. Second, from the base overlay UMM extracts *source-specific multicast distribution trees*: subsets of tunnels in the base overlay used to route and deliver messages generated by a specific source.

UMM starts with a random base overlay and uses a 'short-long' heuristic to incrementally improve it (Section IVB.). Additionally, the base overlay provides the basic bootstrap, fault-recovery, and connectivity maintenance functionality (Sections IVC to IVE). Consistent with our goal of layer independence, the heuristics and mechanisms used at the base overlay layer are independent of the mechanisms employed by higher layers.

UMM builds source-specific dissemination trees from the set of tunnels offered by the base overlay in the following way (Section IVA): a new multicast source starts by flooding the base overlay. Participating nodes observe this traffic and use the implicit information contained in duplicated messages to filter-out the redundant, low-quality tunnels that generate duplicate traffic.

# III OVERLAY TOPOLOGY CHOICES

In order to place UMM in context, this section presents a minimal overview of overlay topologies used to support multi-source multicast while Section VII surveys related work in depth. Depending on the topology of the supporting overlay, approaches to providing multicast functionality can be grouped in three broad classes: shared-tree, unstructured, and structured overlays.

## *A. Shared-tree Overlays*

Shared-tree overlays maintain a tree-structured topology [5, 9, 21] and generally target single-source multicast. The tree topology makes explicit routing mechanisms unnecessary: a node simply forwards each new message on all its tunnels except the tunnel from which it received the message. Shared-tree structures have major limitations though: trees are fragile with respect to node failures, introduce additional delay penalties, and do not make effective use of all bandwidth available in the underlying network [22].

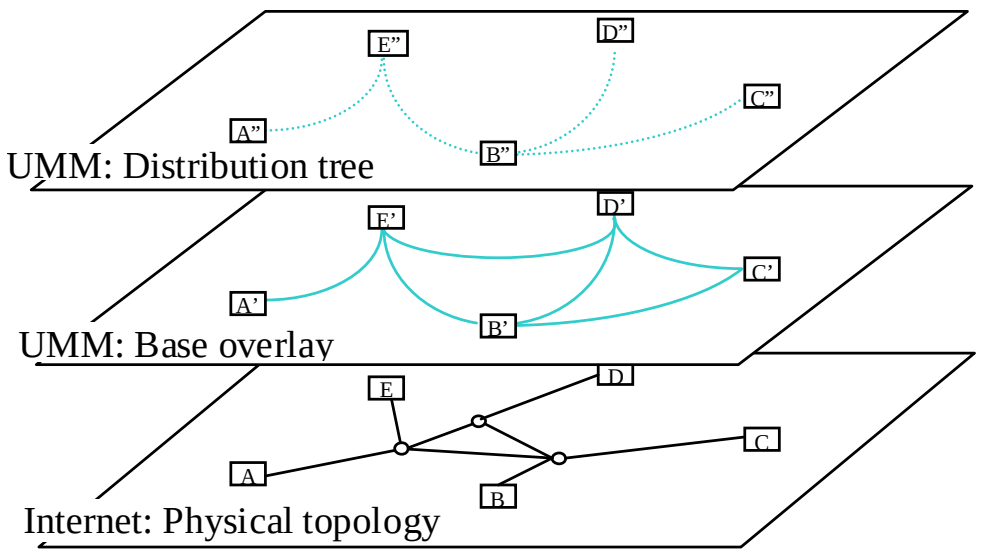

**Figure 1:** Layered view of UMM - five nodes are linked in a physical topology, a base overlay, and a distribution tree rooted at node B.

### *B. Mesh-based, Unstructured Overlays*

Narada [8] and Scattercast [12] build an unstructured, initially random, base overlay mesh and employ a distance-vector routing protocol (DVRMP [23]) to extract source-specific distribution trees. The routing protocol provides each node with information on *all* other system participants and on the optimal cost and path to reach each of them. The considerable volume of data transferred to build and maintain this large state at each node significantly limits scalability: Narada, for example, is designed for groups of up to a few hundred nodes [8].

However, two main advantages result from this approach: first, routes are optimal for a given cost function and base overlay; and second, once detected, connectivity failures can be fixed easily using the routing information. Additionally, the information collected by the routing protocol is used by heuristics that incrementally improve the base overlay, in a solution that couples two functionally independent layers: dissemination tree extraction and base overlay optimization.

Some peer-to-peer (P2P) applications (e.g., Gnutella) use flooding over a random base overlay to distribute messages. While simplicity is an undeniable advantage of this approach, its main drawback is inefficient network usage due to duplicate messages resulting from ignoring the underlying physical network topology [1].

### *C. Structured Overlays*

The regular overlay structures at the base of distributed hash tables (e.g., hypercube, Plaxton mesh) can be exploited to extract multicast distribution trees. While originally targeted for different purposes, there is now interest for using these structured overlays to support application-layer multicast and wide-area data dissemination [14, 15, 24, 25].

Two important benefits have been asserted for this approach. First, good scalability through reduced state size at each node: structured overlay routing tables grow only logarithmically with the size of the network. Second, the ability to support multiple multicast groups and reuse the overlay structure. These benefits however, are subject to debate. In a recent paper, Bharambe et al. [16] argue that, under realistic deployment conditions involving heterogeneous end-host and link capacities, these benefits are drastically limited: optimized dissemination trees that achieve good performance must include a large number of tunnels that are not part of the original structured overlay, thus increasing the state maintained at each node. Additionally, these tunnels limit the benefits of route convergence and loop-free properties of the original structured overlay and generate additional maintenance costs, thus limiting scalability.

## IV UMM DESIGN

This section presents in detail UMM's two functional layers as well as the techniques used to deal with node failure, adapt to network dynamics, and bootstrap the network.

### *A. The Base Overlay*

UMM uses relatively simple heuristics to build and maintain the base overlay: the aim is simply to include efficient distribution trees that can be later selected by the upper layer. This section presents a high level view of the base overlay management operations while Appendix XA presents the pseudocode.

While the base overlay is initially random, nodes optimize it incrementally, an approach common to other systems [8, 17, 26]. In keeping with our decentralization and self-organization design principles, nodes decide independently based only on local information what overlay tunnels to add or delete. As in Shen et al. [17], nodes maintain a fixed proportion $P_{short}$ (defaulting at 50%) of their tunnels *short*, i.e., to nearby nodes, and the other tunnels *long*, to distant nodes. We use a latency threshold $D_{short}$ (with a 10ms default value) to discriminate between short and long tunnels. At fixed time intervals, a node runs an optimizer task that randomly selects a small subset of participating nodes, evaluates new tunnels, and replaces its worst tunnel with a new one, if better. We limit changes to one tunnel replacement per optimizer iteration.

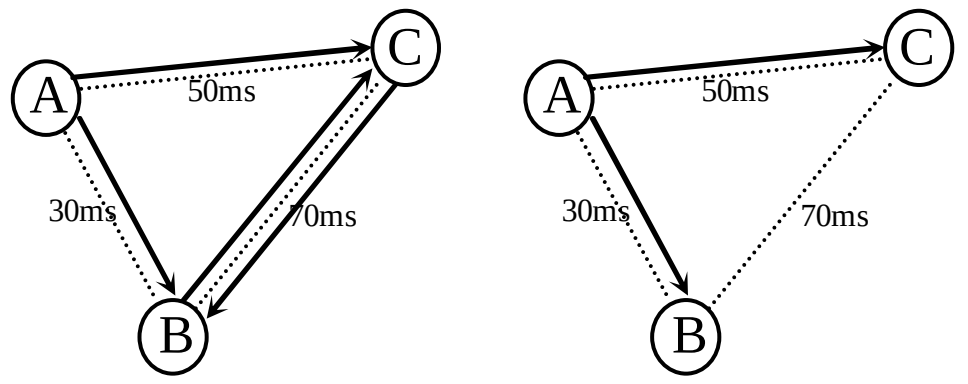

**Figure 2:** Extracting a dissemination tree rooted at A. Left: Flooded messages from source A before B and C detect that tunnel BC is unnecessary. Right: B and C filter out the link BC for messages sourced at A.

Short tunnels are optimized for latency: that is, a node replaces its worst short tunnel if a new shorter tunnel is found. Long tunnels are optimized for bandwidth: that is, a node replaces its worst long tunnel if it finds a new tunnel with better bandwidth. To avoid oscillations due to measurement errors or variations in network conditions, a threshold is used when deciding whether to replace an existing tunnel.

In addition, the overlay topology can adapt to heterogeneous node characteristics by defining the number of tunnels a node supports, and thus its load, to be proportional to node capacity (e.g., its access link bandwidth).

### *B. Extracting Efficient Dissemination Trees*

UMM extracts efficient, source-specific dissemination trees using the implicit information from duplicated messages resulting from flooding. The key idea is that an initial flooding stage allows each node to infer properties of the base overlay just by detecting duplicate messages: duplicates arrive on paths with worse properties than those taken by the first message. Once duplicates are detected, this information can be used to filter out redundant paths for efficient message dissemination. The rest of this section presents an informative example while the Appendix XB presents the data structures and the pseudocode for the protocol executed at each node.

In Figure 2 (left), nodes A, B, and C form an overlay. A is the source of a multicast message $m_A$ that is flooded to B and C. B and C in turn flood the message on all their other tunnels. Nodes that receive a duplicate do not forward it further. Assume B receives $m_A$ first from A and then from C. B can thus infer that it has a better path from A than the one that goes through C and can ask C not to forward further A-sourced messages on the C-B tunnel. The same mechanism, when used at C, causes B to be asked to stop forwarding A-sourced messages to C on the B-C link. Thus, nodes B (respectively C) install 'filters' for A-sourced messages for their tunnels to C (respectively B), and a distribution tree as in Figure 2 (right) is extracted using only the duplicate message information that flows into the network. Note that the distribution tree obtained with this mechanism is cost efficient with respect to the costs of sending A-sourced messages.

### *C. Dealing with Node and Tunnel Failure*

Node and tunnel failures are the main potential problem once dissemination trees have been built by filtering out redundant tunnels as presented in the previous section.

Failure detection is straightforward: nodes paired by tunnels periodically exchange aliveness messages. If one node does not receive these messages for a predefined period of time, it infers that a failure occurred involving either the other end of the tunnel or the underlying network path.

Once a failure is detected the node initiates the recovery process. First, to restore the connectivity of distribution trees that used the tunnel, the node that detects the tunnel failure resets all filters for tunnels ending at the node. Second, nodes must also deal with tunnel failures that occur further away in the overlay topology, to prevent local failures from destroying connectivity to remote nodes. To address this latter requirement, nodes that detect a failure flood a *resetRouteMessage* message with a small time-to-live that causes all receivers to reset their 'filters' and to eventually restart the process of tunnel selection described in the previous section. The corresponding pseudocode is presented in the Appendix XB.

Furthermore, filters are soft-state: they expire after a certain timeout. Thus, nodes regularly revisit their decision to include or exclude a tunnel from a distribution tree. Timeouts can be fixed parameters set based on application-specific knowledge, or can be computed adaptively, in a manner inspired by TCP timeouts. A related approach is employed in other P2P systems [1, 27] based on the intuition that the 'age' of a node is a good predictor for its future lifetime.

### *D. Keeping the Base Overlay Connected*

UMM splits this task into two separate problems: (1) partition detection and (2) repair.

*Partition detection*. As in [12, 17] to detect partitions we employ a heartbeat-based solution. A special heartbeat node that inserts messages in the network at regular intervals $T_{htbeat}$. Nodes that do not receive the heartbeat for some multiple of $T_{hbeat}$ conclude that they have been disconnected and initiate the repair procedure. If the heartbeat monitor dies, Saxons [17] suggests an election protocol to elect a new heartbeat source. As this protocol is unlikely to work at our target scale, we use $N$ heartbeat sources: a node is connected as long as it hears from more than $^{N}/_{2}$ sources. This solution tolerates $^{N}/_{2}$-1 simultaneous heartbeat source failures.

*Repair*. Once a node detects it is disconnected it first flushes the set of other nodes it knows about, and tries to rejoin the network, starting from the heartbeat nodes it did not hear from or from a bootstrap node. To avoid flooding the bootstrap node when multiple nodes detect they have been disconnected simultaneously, a node waits for a random time before initiating the repair procedure.

An important optimization is preventing partitions induced by the continuous adaptation of the base overlay. To this end, nodes avoid dropping the tunnels over which they receive heartbeats. Similarly, when a node drops a tunnel to add a new neighbor, it makes sure the new neighbor does not receive heartbeats through the tunnel that it planned to drop.

### *E. Membership Maintenance and Bootstrap*

Each node maintains a random subset of identifiers of other nodes participating in the overlay. UMM employs an epidemic style solution to maintain this membership information: two arbitrary nodes contact each other, for example to decide whether they agree to establish a new tunnel; they also exchange a small random subset of identifiers of other participants in the overlay. This solution, akin to epidemic communication of membership information, results in uniform distribution of this information across the overlay, has scaled well in existing deployments (e.g., Gnutella [1] has reached more than one million nodes), and has been studied extensively [12, 17].

To join the overlay, a new node needs to be able to contact one node participating in the network. The joining node contacts the already participating node aiming to establish a new tunnel. Even if the contacted node does not have available incoming slots, it will return a set of addresses of other participants thus enabling the joining node to continue its search for nodes to connect to. In the current implementation, all nodes know the address of a fixed bootstrap node, but alternative solutions that do not rely on fixed, predefined bootstrap node are possible [28].

## V IMPLEMENTATION

We have implemented UMM in Java. The implementation includes about 25 main application classes and about 60 auxiliary classes for a total of 5,055 lines of code.

To use UMM, an application uses a *send_message* non-blocking call that puts messages in the appropriate outgoing message queues at the local node. The application also needs to register a callback method that is called when messages addressed to the application arrive. The application can also register callbacks to be notified when the base overlay or the dissemination tree change.

The transport protocols used to implement UMM tunnels need to provide two properties: TCP friendliness and ability to express reliability characteristics in terms of application level frames. For convenience, in the experiments presented in this section we use TCP as a base for UMM transport protocol. We are now currently experimenting a new, UDP-based transport protocol, that provides framing, implements TFRC [29] for congestion control, and allows application to control the tradeoff between transport reliability and timely frame propagation.

To estimate available bandwidth and path latencies, we have selected *netperf* [30] from a large number of network measurement tools (Spruce, PathLoad, TOPP and others) for its ability to perform well when deployed on PlanetLab. Each host runs a *netperf* server along with UMM. We modified the *netperf* client to work as a library that UMM invokes through Java Native Interface.

## VI EXPERIMENTAL EVALUATION

This section presents experiments that evaluate UMM techniques described above and the effectiveness of our UMM implementation: First, in Section VI.A we use an emulated network (ModelNet [20]) to perform extensive controlled experiments that allow us to compare UMM with IP-layer multicast and with alternative solutions using relative delay penalty and network stress metrics. Second, in Section VI.B we explore UMM performance under real-world conditions when deployed on PlanetLab [19], a live network testbed.

For the experiments presented in this section, nodes are configured to initiate at most five tunnels and accept at most seven, thus the average node degree is ten. Half of these tunnels are configured as *short*, thus optimized for delay, and half *long* and optimized for bandwidth. The timeout for a tunnel filter is fixed at 600s. The tunnel optimization thresholds mentioned in Section IVB are set at their default values: 2 ms for delay and 50% bandwidth improvement. While Section VI.A.5. presents a simulation methodology and results for setting tunnel optimization thresholds, for the other parameters we are currently performing extensive sensitivity studies to gauge their impact on the overall performance of the system.

### *A. Experiments in an Emulated Environment*

ModelNet [20] is an emulation environment for wide-area networks. The target application runs unmodified on a set of cluster nodes. ModelNet extracts path delay and bottleneck bandwidth from user provided network topologies and emulates wide-area network traffic conditions for communication among these cluster nodes. In order to achieve scalability, ModelNet offers tunable emulation fidelity: depending on the desired fidelity level, packet contention on shared physical links is not emulated at border routers.

We experiment with UMM in this controlled environment with *three goals* in mind. First, we compare the quality of the dissemination trees produced by UMM with that of trees built by alternative approaches. Second, we estimate the cost of self-organization: we benefit from having full information on the network topology and we compare UMM-generated dissemination trees with optimal trees generated by heuristics using global topology knowledge. Finally, we test UMM ability to operate and adapt when the set of participating end-hosts is dynamic.

*1.) Experimental Setup*

We use the ModelNet suite of tools to generate physical network topologies: ModelNet in turn uses Brite [31] topology generator to generate the topology graph and assign link latencies. Modelnet classifies network links as Client-Stub, Stub-Stub, Transit-Stub, and Transit-Transit depending on their location in the network [22]. Bandwidth for each link is assigned depending on its type: Client-Stub links have bandwidth randomly assigned form the 2 to 8Mbps range, Stub-Stub links: 4 to 10Mbps, Stub-Transit links: 5 to 10Mbps, and Transit-Transit in the 10 to 20Mbps range.

End-hosts are attached randomly to stub level routers. Overlay nodes then are picked randomly from these hosts. In the experiments presented in this section, for each overlay size we ran ten simulations on ten different physical network topologies. Each topology has 4,040 backbone routers. Simulated overlay size varies from 64 to 1,024 nodes.

*2.) Relative Delay Penalty and Network Stress Comparison*

This sub-section, starts by describing techniques to estimate relative delay penalty (RDP) then compares the quality (in terms of RDP and network stress) of UMM produced dissemination trees with that of trees built by implementations of the alternative approaches surveyed in Section III.

There are two approaches for estimating propagation delays, thus the RDP, in the overlay.

- The *offline* approach extracts the map of the overlay and combines it with network topology information when using Modelnet (or with delay estimates when on PlanetLab) to compute, offline, the delay for each overlay path.
- The *online* approach estimates propagation delays through the overlay itself: once the distribution trees are extracted: a node multicasts a special *ping* message to which each participating node replies. Replies are propagated back to the source through the overlay. Delay estimates using this method are inherently more pessimistic: the source and intermediary nodes need to process, in sequence, a potentially large number of replies.

For ModelNet experiments these two approaches yield similar results. We discuss the differences we observe on PlanetLab between RDP estimations using these two approaches in Section VII.B.

To compare the quality of the dissemination trees produced by UMM with that of trees built by alternative approaches, we use two sets of comparisons: First, we perform a head-to-head comparison of UMM with Macedon-generated [32] implementations of alternative approaches running on the same emulation platform. Second, we compare UMM performance with that of alternative approaches as reported in a previous simulation-based study by Jain et. al [33]. In this latter comparison, we use ModelNet and attempt to reproduce the physical topologies used by Jain [33] in order to offer a meaningful comparison base.

Macedon [32] is a code-generation tool for overlays: Macedon generates overlay implementations from high-level functional specification of overlay behavior. In its latest version (v1.2.1), Macedon offers specifications for multicast overlays based on shared trees (a random tree and a solution based on Overcast [5] protocol) and on structured overlays (two variants of the Scribe protocol [14]). We have instrumented the generated code to extract the multicast dissemination trees these solutions build at runtime and we compare them with trees built by UMM.

Figures 3 and 4 summarize the results of this comparison for overlays varying from 64 to 1024 nodes. These results show that UMM performs better than the alternative solutions mentioned above. In all experiments the *90%-tile RDP offered by UMM is lower* (thus better) than that offered by alternative solutions (Figure 3). On average, the 90%-tile RDP is 29% better than structured overlay solutions based on Scribe, 58% better than for random trees, and more than five times better than for Overcast. We note that the two versions of Scribe available (implementing optimizations proposed by groups at Rice University and Microsoft respectively) performed similarly. We were unable to run Scribe with more than 512 nodes due to a file descriptor leak in the Macedon implementation.

In part due to our configuration to use similar fan-outs for the dissemination trees, all solutions we test offer comparable *maximum link stress* (Figure 4). The exception is the random trees which present a much more accentuated growth of maximum link stress for large overlays. Note that, although Overcast appears to produce a slower increase in link stress for large overlays, this result is achieved using a distributed lock to sequence network measurements, a technique that is clearly non-scalable. (In fact, the time to bootstrap a 1,024 node Overcast overlay grows as large as six hours.) Without using the distributed lock mechanism the link stress evolution is in line with that produced by UMM and Scribe.

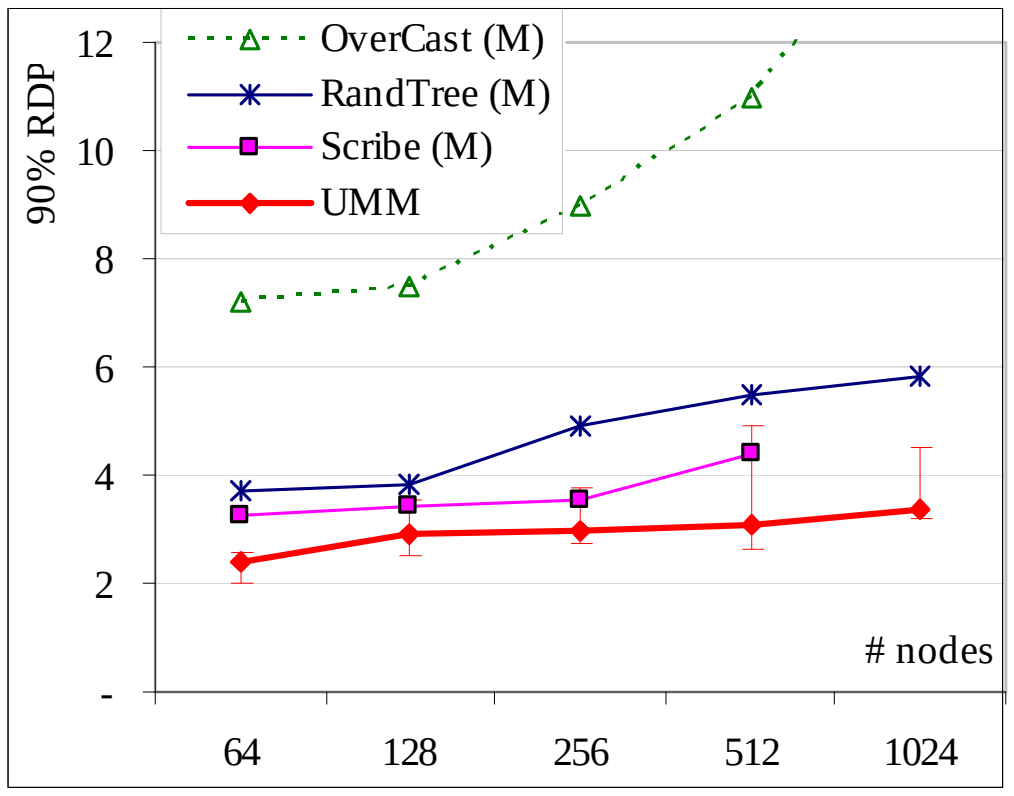

**Figure 3**: 90%-tile RDP for UMM and alternative solutions for 64 to 1024 node overlays. Each point presents, average values from at least 10 runs over different physical topologies. M stands for Macedon implementation.

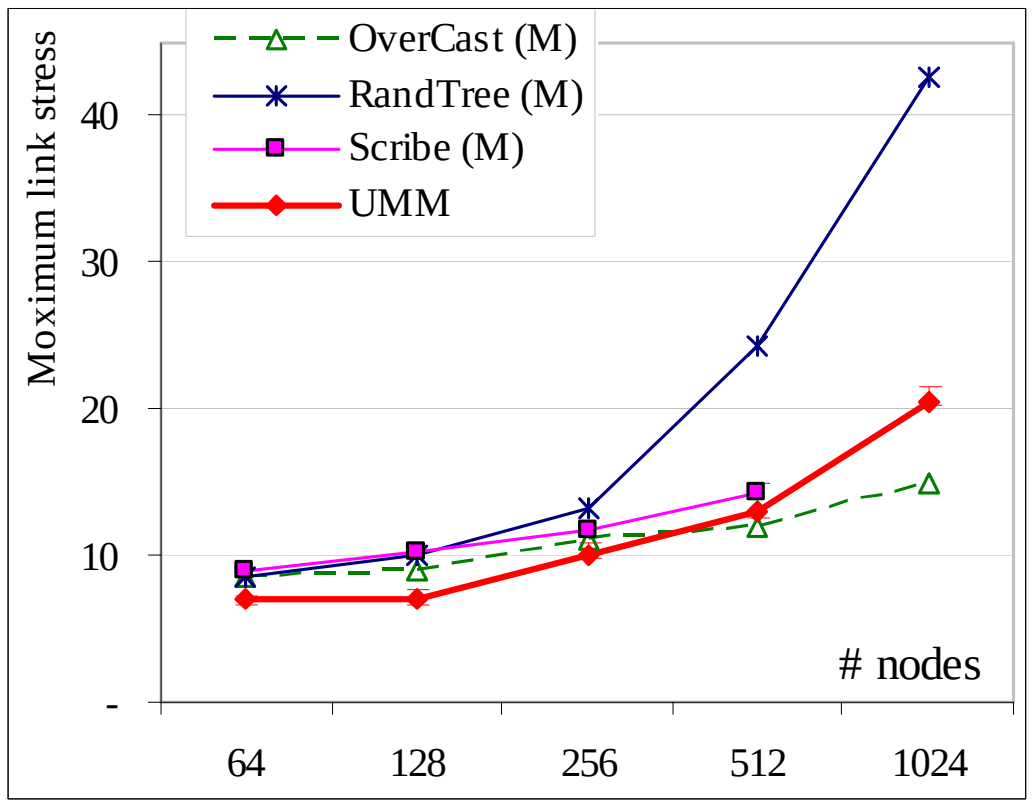

**Figure 4**: Maximum link stress for UMM and alternative solutions for 64 to 1024 node overlays. Each point presents average values from at least 10 runs (20 for UMM), while the error bars for UMM present min. and max. values.

The second set of experiments compares UMM performance with that of alternative approaches reported by Jain et al. [33] in a previous study. Jain considered three unstructured overlays (Narada [8], Nice [6], and a power-law overlay that is mapped randomly on the physical topology) and structured overlays employing various mapping heuristics.

We attempt to reproduce the physical topologies used in Jain study to offer a sound comparison base. Figure 5 and 6 summarize the results of this comparison for overlays varying from 64 to 1024 nodes. These results show that UMM generally performs at least as well as other alternatives: The 90%-tile RDP offered by UMM is better than that obtained using *realistic* structured overlays (i.e., structured overlays that do not assume a global view when mapping on the physical network). Compared to solutions that assume global view for optimization (such as Narada and the *idealized* structured overlays), the *90%-tile RDP* offered by UMM is slightly higher, thus worse (Figure 5). In all experiments UMM offers lower (thus better) *maximum link stress* than Narada, structured overlays, and naïve flooding on power-law random graphs (Figure 6).

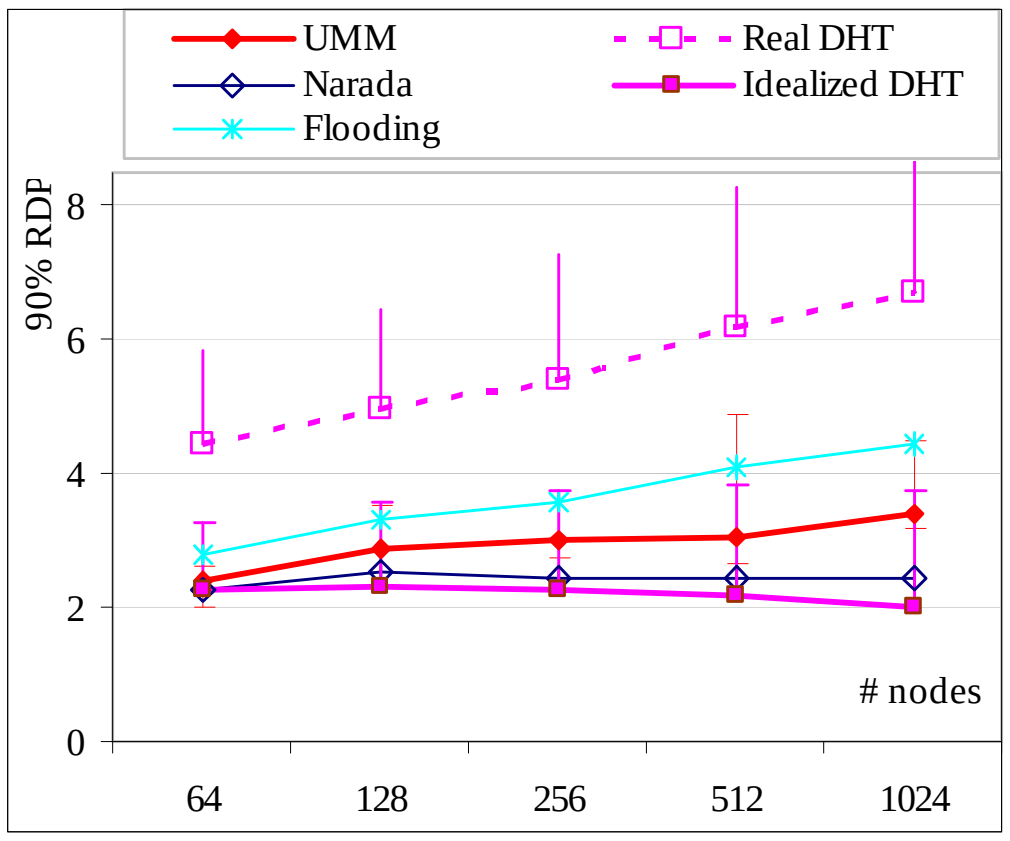

**Figure 5**: 90%-tile RDP for UMM and alternative solutions for overlay size varying from 64 to 1024 nodes. For UMM, Narada, and flooding, each point presents, average values from at least 9 runs (20 for UMM), while error bars present minimum and maximum values. Structured overlay results (labeled as DHT in the plot) present the average RDP for the best of three heuristics considered by Jain while the performance of alternative heuristics is presented in the error bars. Non-UMM results are extracted from Jain et al. [33].

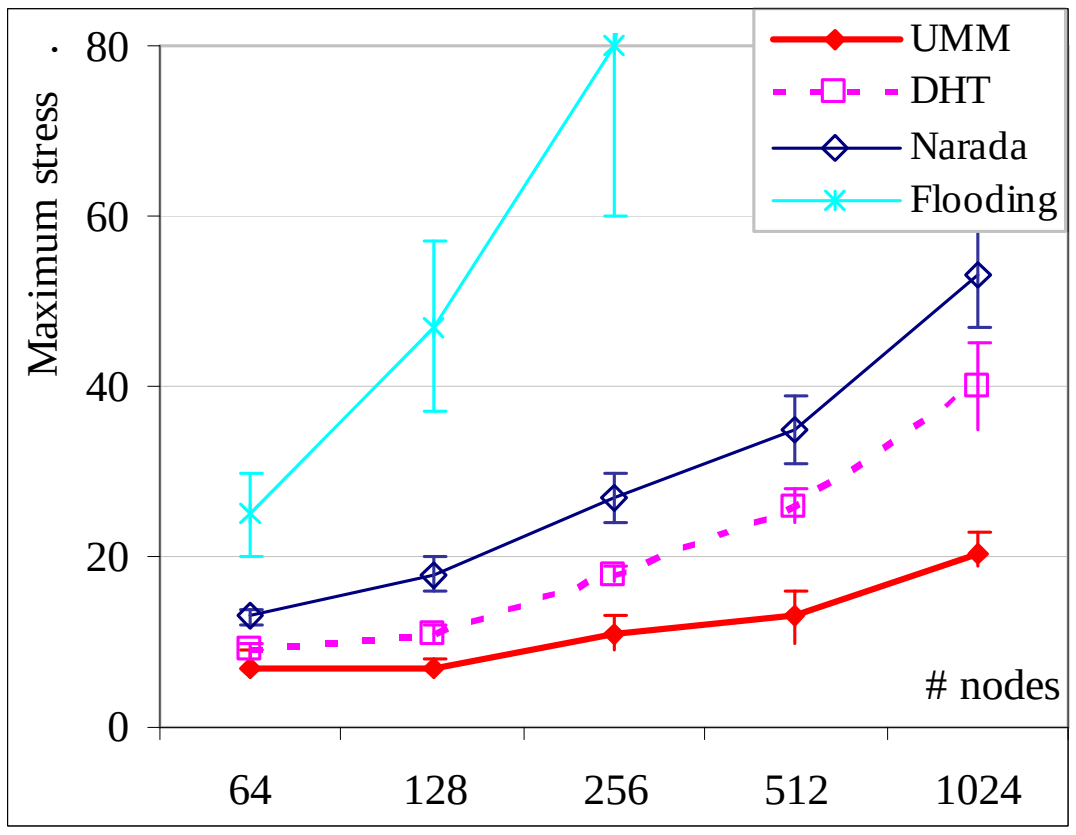

**Figure 6**: Maximum link stress for UMM and alternative solutions. Overlay sizes vary from 64 to 1024 nodes. Each point presents average values from at least 9 runs (20 for UMM), while the error bars present minimum and maximum values. Non-UMM results are extracted from Jain et al. [33].

For clarity, Figures 5 and 6 do not reproduce Nice [6] performance results reported by Jain et al. Jain's results show that, compared to Narada, Nice incurs 22% to 60% higher overheads in terms of RDP and performs similarly in terms of network stress. Thus Nice performs similarly to UMM in terms of RDP and worse in terms of network stress.

We note that the Scribe performance observed in our experiments is slightly better than that predicted by the decentralized heuristics investigated by Jain (but still much worse than that of idealized heuristics using global knowledge). We attribute this situation to two reasons: first, Scribe enhances the structured overlay with 'shortcuts' to improve the characteristics of the dissemination trees extracted, a heuristic not explored by Jain et al. [33]. Second, the heuristics to improve overlay mapping and message routing in structured overlays have received significant attention since Jain et al. [33] was published.

We note that Castro et al. [34] concludes that dissemination trees based on Pastry (the structured overlay Scribe uses) offer better performance in terms of relative delay penalty and network stress than solutions based on M-CAN [24]. Since UMM performs better than Scribe, Castro's study gives us confidence that UMM performs better than a larger class of structured overlay based solutions.

***Summary*:** The experiments we have presented support two conclusions: First, we show that it is possible to passive data collection and local decisions to extract *efficient* multicast data dissemination from unstructured overlays. Second, we demonstrate that although UMM uses significantly simpler heuristics than those used by alternative solutions, the data dissemination trees extracted have better or comparable performance.

*3.) Cost of Self-Organization*

This section attempts to answer two intertwined questions:

First, we attempt to estimate the cost of self-organization: How do dissemination trees built by the self-organizing UMM solution compare with optimal trees extracted using centralized heuristics that use full knowledge about the physical topology? The performance of these optimal trees serves as upper bound for the achievable performance.

Second, we attempt to understand where to direct future effort to improve UMM heuristics: Which of the two UMM layers, the base overlay layer or the tree extraction layer, introduce the largest efficiency losses?

In order to answer these questions, we compute, using full topological information, hypothetical dissemination trees optimized to minimize relative delay penalty as described below. We note that computing these trees are variations of the Steiner tree problem [35], and, although we have not proven it formally, we believe that algorithms that find the optimal solutions are NP-complete. In the following, by 'optimal' solution we mean the best solution discovered by the polynomial-time approximation heuristic we describe.

- *The optimal spanning tree that can be extracted from the base overlay UMM builds.* We start from base overlay topologies captured during UMM runs and use Dijkstra's shortest path algorithm to compute optimal dissemination trees with respect to propagation delays. (The corresponding plots are labeled **BASE** in Figure 7).
- *The optimal spanning tree that can be extracted from a base overlay whose construction has also been optimized using global knowledge heuristics*. The base overlay is built using the following heuristic: initially all nodes add *maxdegree* tunnels to their nearest neighbors in terms of delay; if the graph is not connected after this step, the heuristic iteratively picks one node at random, deletes its worst tunnel and adds one to the next closest neighbor in terms of delay until the graph becomes connected. On top of this base overlay dissemination trees are built using Dijkstra. (The corresponding plots are labeled **BESTBASE** in Figure 7).
- *The optimal maxdegree out-degree limited spanning trees built on top of the complete graph*. This heuristic starts from a designated root node and builds a shared tree in a greedy fashion by adding, at each step, the shortest possible tunnel that expands the tree. The only difference from the minimum spanning tree algorithm is that node degree is limited to *maxdegree*. Note that this heuristic will always produce trees with better characteristics than those extracted from a *maxdegree*-limited base overlay. This reflects the difference between single source and multi-source multicast overlays under the same maximum node degree constraint. (Plots labeled **OPT** in Figure 7.)

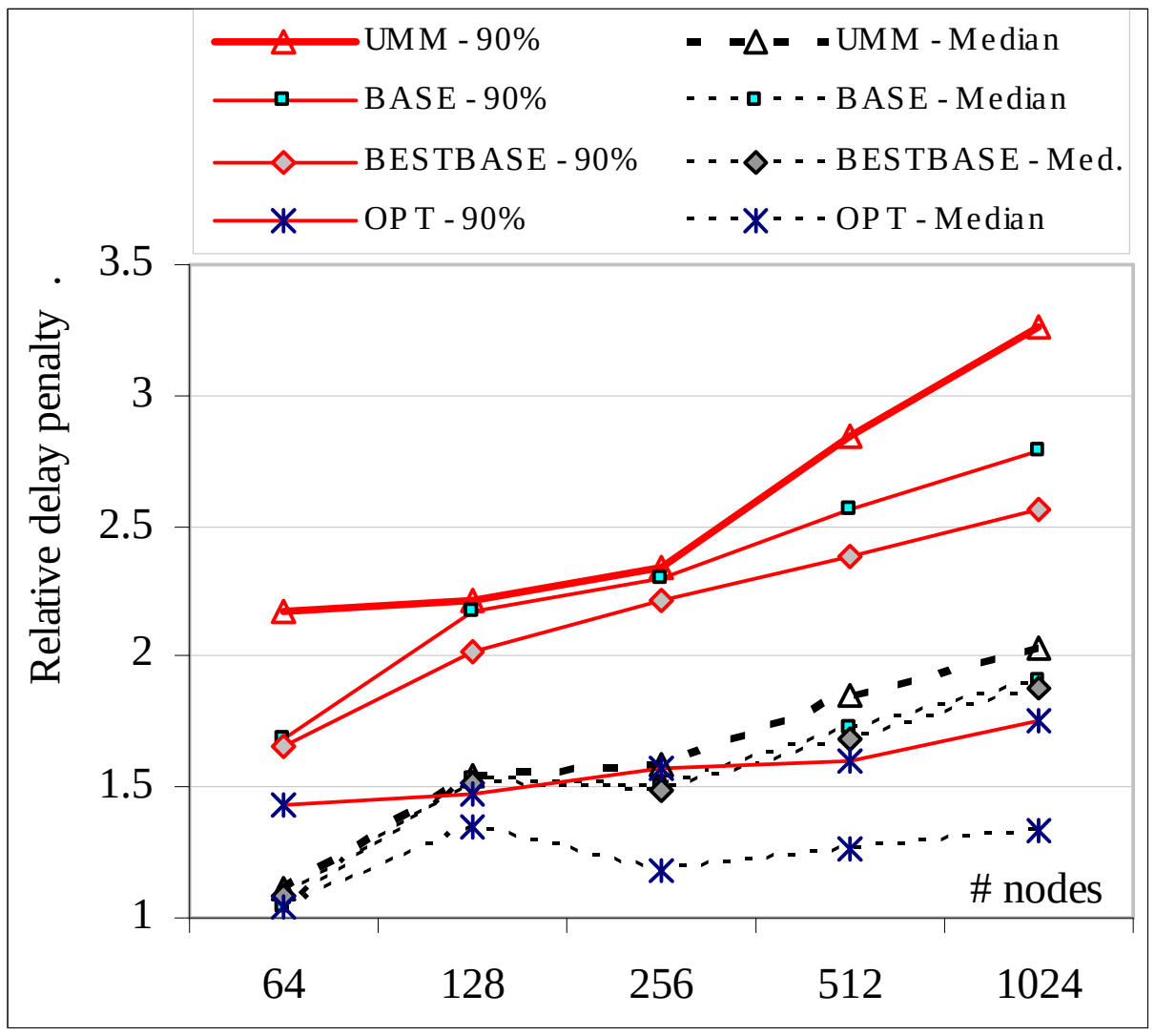

**Figure 7:** 90%-tile (solid lines) and median (dashed lines RDP comparison with distribution trees extracted by heuristics using global knowledge

Figure 7 presents the average 90%-tile and median RDP for disseminations trees built using UMM on ModelNet and using the three heuristics described above for 20 multicast sources for each of the five overlay sizes. These experiments present the distribution of performance loss generated by the two UMM layers. The yardstick is the performance of centralized, global-knowledge heuristics. Averaged over the five overlay sizes we experiment with, the 90%-itle RDP of UMM is 15.1% less than that of the optimal trees extracted from an optimal base overlay. This performance degradation has two components: 5.3% is due to the non-ideal base overlay and 9.8% is due to the non-ideal dissemination tree. In terms of median RDP however, the total performance loss is only 5.4%, almost entirely localized at the dissemination tree extraction layer. A second observation is that, as expected, dissemination trees optimized for a single source (OPT) offer significantly better performance in terms of RDP.

**Summary:** These experiments support two conclusions: First, UMM has low overheads when compared to ideal solutions that use heuristics based on global views of overlay membership and network properties, thus they support our claim that it is possible to use flooding and passive data collection to extract *efficient* multicast data dissemination from unstructured overlays. Second, two thirds of these overheads are concentrated at the dissemination tree extraction layer.

*4.) UMM Under Churn*

An important characteristic of any distributed system designed to be deployed on a real-world platform is its ability to operate in spite of component failures. In this section, we evaluate UMM's ability to deliver messages under 'churn': the continuous process of node arrival and departure.

The performance metric we employ is the ratio of messages actually delivered to the number of messages that ideally should have been delivered under churn. In order to compute the number of messages that should have been delivered under churn, we keep track of all node join/crash events as well as all message generation events. For each receiver *R* and for each message *m* generated by source *S* at time $T_{S(m)}$ we estimate that the message should have been delivered at *R* if the node joined the network before the generation time $T_{S(m)}$ and left the network after $T_{S(m)}+T_{prop}$ where $T_{prop}$ is an estimated upper bound for overlay propagation time (set conservatively at 5s in the experiments we report here). In other words, node *R* should receive a message issued at time $T_{S(m)}$ if: $T_{R(join)} < T_{S(m)} < T_{R(leave)} - T_{prop}$.

To model churn, we use independent node failures controlled by a Poisson process. This model is based on studies of user behavior in multicast groups on the MBone [36] and in file sharing applications [37] and has been extensively employed [26, 38, 39]. As a result of the Poison process, node lifetimes are exponentially distributed. In experiments, we vary the median node lifetime from 300s to 7200s to infinity (no failures). Studies of various file-sharing networks report median node lifetimes between 10 minutes [40] and one hour [37, 41]. In our experiments, a new node joins the overlay 10s after a node failure to keep the size of the overlay constant.

Consistent with our aim to gauge UMM behavior in an unfriendly environment, we model crash-like failures as opposed to graceful (announced) node leaves. This inherently leads to higher loss rates: unlike for announced leaves, where it is possible to flush node buffers and exit gracefully, under a crash model messages are lost when nodes crash with non-empty queues.

Figure 8 presents the results of three sets of experiments with 128, 256 and 512 nodes, respectively, running on ModelNet. Each node inserts small messages into the network at a rate of one message every 5 second (so the system will deliver up to 52k messages/second). We use small messages as we attempt to evaluate the ability to timely detect node crashes, repair the overlay structure, and maintain the overlay connected, while we are less interested here in the ability to evaluate the transport capacity.

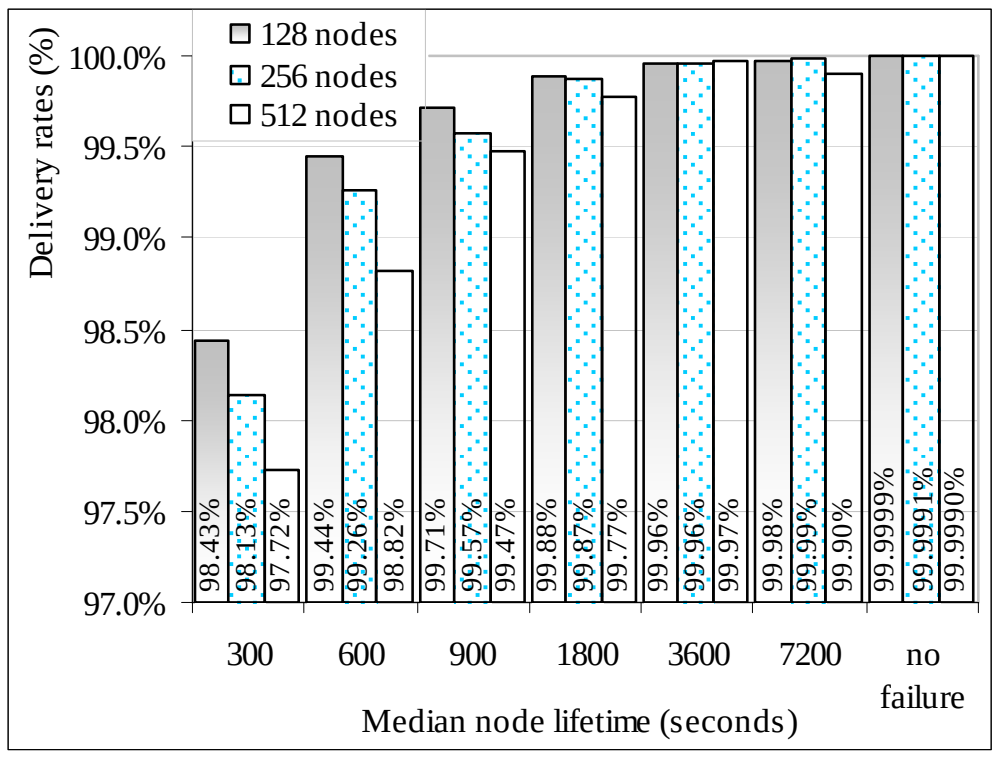

**Figure 8**: UMM delivery rates under churn. Average delivery rates for 128, 256, and 512 nodes overlays under various mean node lifetime assumptions.

As Figure 8 shows, for 512 nodes with 5-minute average lifetime (implying perhaps unrealistically high failure rates of 1.7 failures per second on average in the whole system) 97.72% messages are delivered on average. Under moderate failure rates of 1-hour average node lifetime (translating, to one failure every 7s on average in a 512 nodes system) more than 99.9% messages are delivered on average over the whole experiment. With no node crashes, fewer than one message in 100,000 is lost.

How do these results compare with the performance of other systems? We attempted to evaluate, under the same scenario, the performance of the three other overlay multicast solutions provided by Macedon and mentioned in the previous section. Unsurprisingly, the random tree solution performs fairly well as the cost of restructuring the overlay after a node failure/join is minimal: the failure rate stays above 99% for all experiments with 512 nodes. However, Macedon implementations of Overcast and Scribe do not handle failures: they become unusable even under low failure rates. (We believe this problem is an implementation shortcoming and not an inherent characteristic of these classes of solutions).

***Summary*:** Although UMM does not include explicit resiliency mechanisms, it is able to recover quickly from node failures and performs well under churn.

*5.) Agility vs. Stability: Bootstrap Convergence*

The ability to converge quickly to a stable state with good properties is an important characteristic of systems that operate in a dynamic environment. In this section, we assume the characteristics of the underlying network are stable and we investigate base overlay convergence after bootstrap. First, as a consequence of the greedy base overlay construction strategy where each node, at each time-step, improves its existing tunnels, the base overlay will always converge to a stable state when no external events occur (i.e., no node failures and no variation in underlying network characteristics).

Second, we examine the tradeoff between convergence time and the quality of the stable state to which the base overlay converges. The variables that determine the shape of this tradeoff space are the thresholds that trigger a tunnel change: As we mention in Section IV.B, a node replaces an existing short tunnel only if the new tunnel offers at least DELAY_THRESHOLD ms. lower delay and replaces a long tunnel only if the new tunnel offers at least BW_THRESHOLD more bandwidth.

**Table 1**: High and low threshold values for tunnel adjustment.

| | DELAY_THRESHOLD | BW_THRESHOLD |
|---|---|---|
| Low threshold | 0.5 ms | 10% |
| High threshold | 8.0 ms | 75% |

Figure 9 and 10 explore the impact of these threshold values on convergence time, number of reconfiguration events, and the quality of the stable state to which the base overlay converges. These figures present averages over 10 runs for a 1024-node overlay for the two extreme sets of threshold values presented in Table 1. At each iteration step, a node probes the network path to ten other overlay nodes, and attempts to replace one short tunnel and one long tunnel. Results are presented in terms of number of iterations, to factor out the duration of a single iteration which is largely determined by the effectiveness of the probing technique.

As expected, higher thresholds for tunnel adjustments lead to faster convergence and a more stable overlay (i.e., fewer reconfiguration events). As Figure 9 shows, for high threshold values, the total rate of reconfiguration events drops to less than one event per time step for the 1024-node overlay after only 10 time-steps, while for low thresholds, the convergence rate is significantly slower: reconfiguration events rate drops to the same level only after 31 time-steps. The total number of reconfiguration events is reduced by more than half when using high thresholds.

The difference in convergence time is also reflected in the characteristics of the stable state obtained using the two configurations above. When using high threshold values, the average tunnel latency is 6.8% worse and the average tunnel bandwidth is 18% worse than when using low thresholds. However, this small degradation in the resulting base overlay quality is acceptable to obtain a more stable base overlay. Moreover, the loss in performance is attenuated when actually extracting dissemination trees from the stable base overlay: 90%-tile RDP is only 4.45% worse.

***Summary*:** We present experiments that correlate stability: i.e., the quality of the base overlay UMM converges to, and agility: i.e., the speed of the convergence process. These results can be used to tune the parameters that determine node sensitivity to changes in the environment to achieve desired levels of service depending on the characteristics of a specific deployment platform.

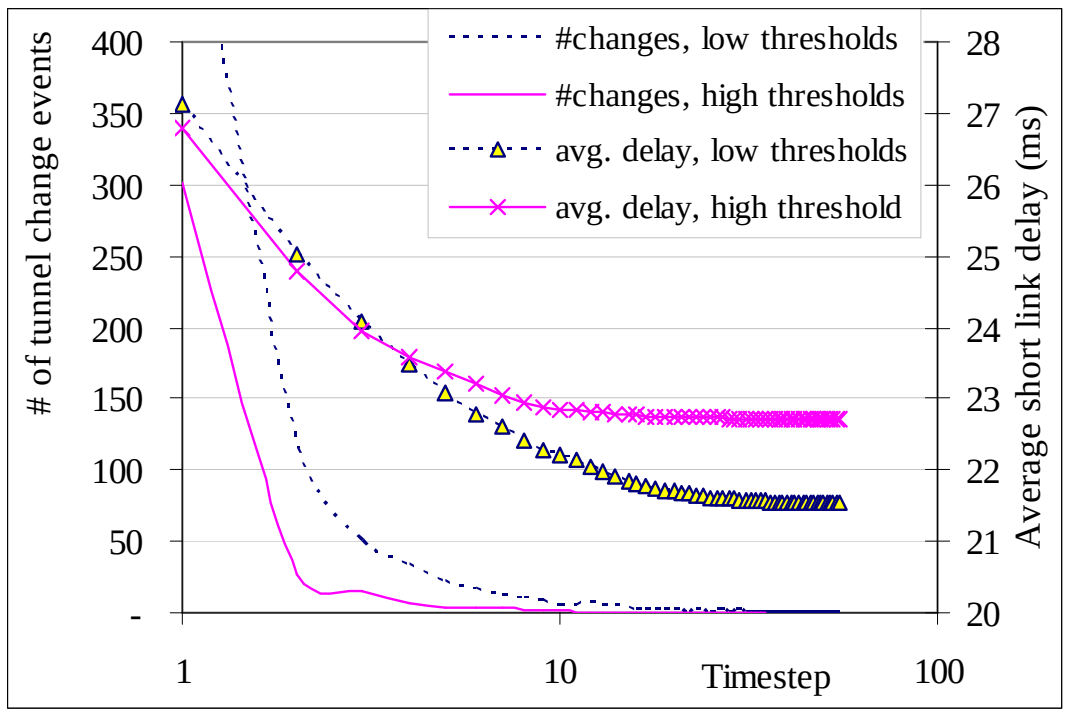

**Figure 9**: Average tunnel delay of short tunnels, convergence time, and number of reconfiguration events and for low and high values of reconfiguration thresholds.

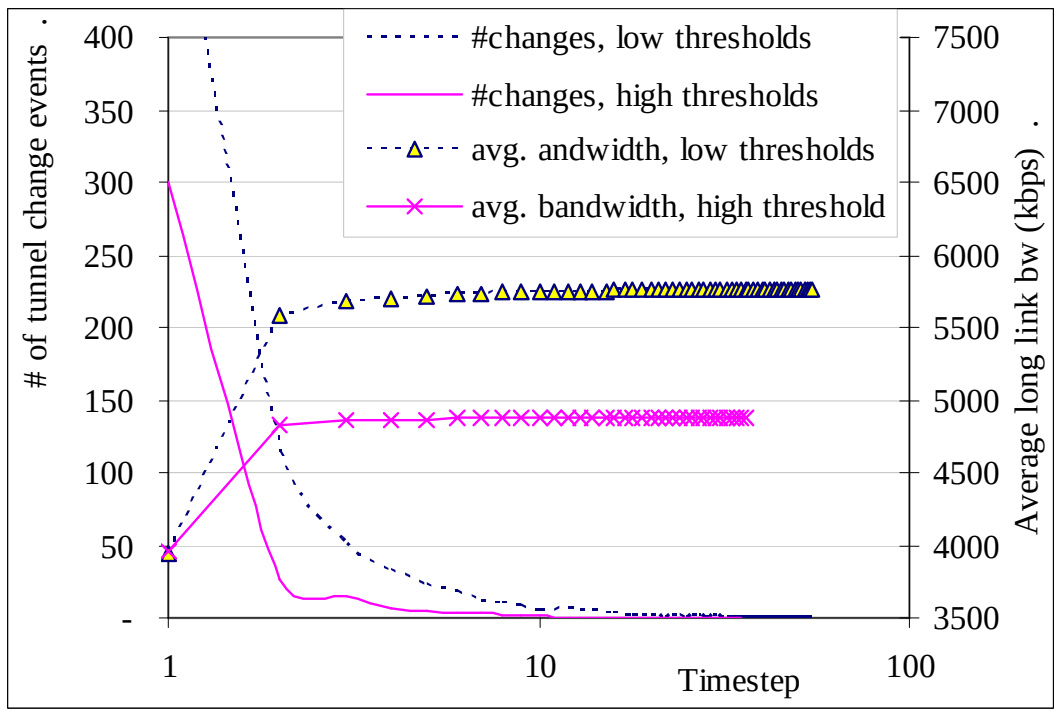

**Figure 10**: Average tunnel bandwidth of long tunnels, convergence time, and number of reconfiguration events and for low and high values of reconfiguration thresholds.

*6.) Summary of ModelNet Experiments*

**Table 2:** The difference between on- and off-line RDP estimates for PlanetLab experiments. For ModelNet the difference is insignificant.

| | Delay estimate | |
|---|---|---|
| | Offline | Online |
| 95%-tile RDP | 2.53 | 4.96 |
| 90%-tile RDP | 1.95 | 3.78 |
| Median RDP | 1.22 | 2.11 |

We have used ModelNet, a wide area network emulator to evaluate UMM performance and compare it with that of alternative approaches. Direct comparisons on ModelNet and comparisons with previously published results support our claim that the dissemination trees produced by UMM have better characteristics (better RDP for the same stress) than those built by alternative approaches (Section VIA.2.). Additionally, the performance degradation brought by the simple UMM heuristics based on *local* information is small when compared with heuristics that employ *global* topological information to extract optimized dissemination trees (Section VIA.3.)). Finally, we have demonstrated the ability of UMM to operate in dynamic environments and to adapt when the set of participating end-hosts is dynamic (Sections VI.A.4 and VI.A.5).

*B. PlanetLab Experiments*

We also ran multiple experiments on PlanetLab [19], a live, wide-area, network testbed. Our aim is to understand how well UMM operates in a real environment. Due to lack of space we limit here to present results from three experiments: first, we evaluate the difference between online and offline RDP estimates when using PlanetLab, second, we present throughput results from a multi-source multicast experiment using 60 PlanetLab nodes, and, third we evaluate agility in dealing with the failure of a sizeable number of network participants when deployed over a live testbed.

*1.) Online and Offline RDP Estimates*

As we mentioned in Section VI.A.2 RDPs can be estimated *online*, by directly measuring the time messages travel in the overlay, or *offline*, by mapping the overlay to point to point delay measurement between nodes. As PlanetLab nodes are heavily loaded (load is often above 10), the delays between the intervals when the application is scheduled can be large, sometimes in the same order of magnitude as IP-layer latencies. Consequently, the RDPs computed using online delay estimates (right column in Table 2) are significantly larger than when computed offline. We have not observed such delays when running on ModelNet, even when 15 copies of our application ran concurrently on the same physical node. Thus, we believe that the larger online RDP estimates are simply a result of high, sometimes transitory, load on PlanetLab nodes.

### 2.) *Throughput on PlanetLab*

Figure 11 presents received data rates at each destination when all 60 nodes participating in the experiment send data at rates varying from 2kbps to 16kbps (resulting in aggregate rates between 120kbps and 960kbps). Nodes are located at distinct sites on three continents. Message size is 4KB for all experiments.

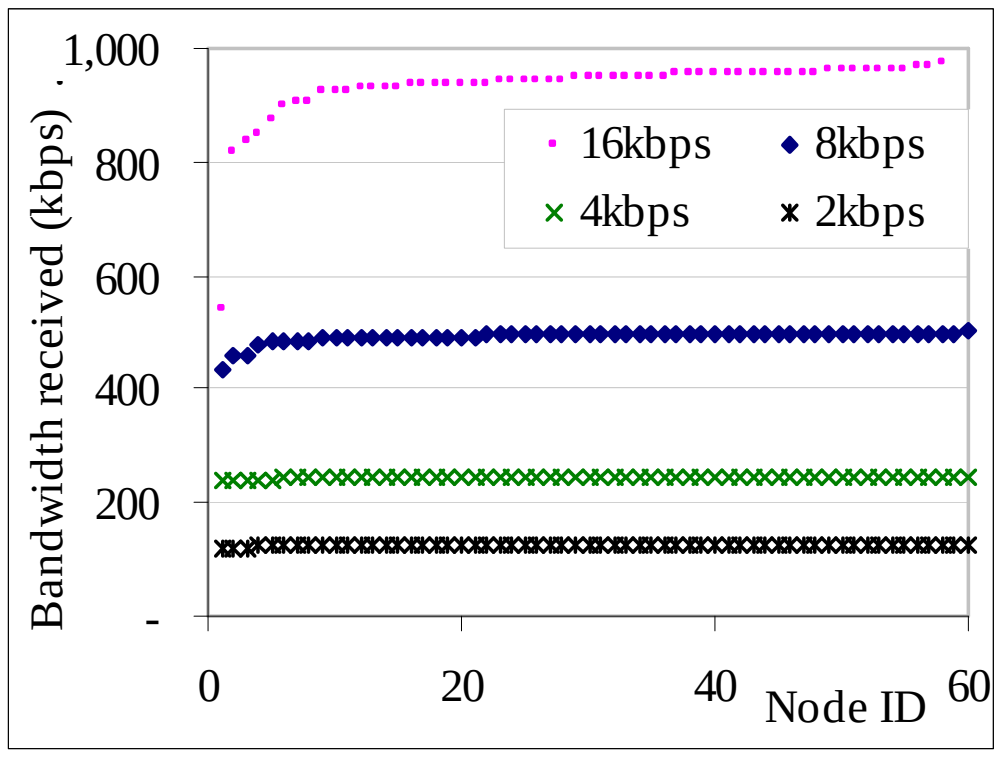

**Figure 11**: Multi-source multicast performance. Each of 60 nodes sends data at rates from 2kbps to 16kbps. The graph presents received data rates at each destination. Nodes are ordered in increasing order of their received rates.

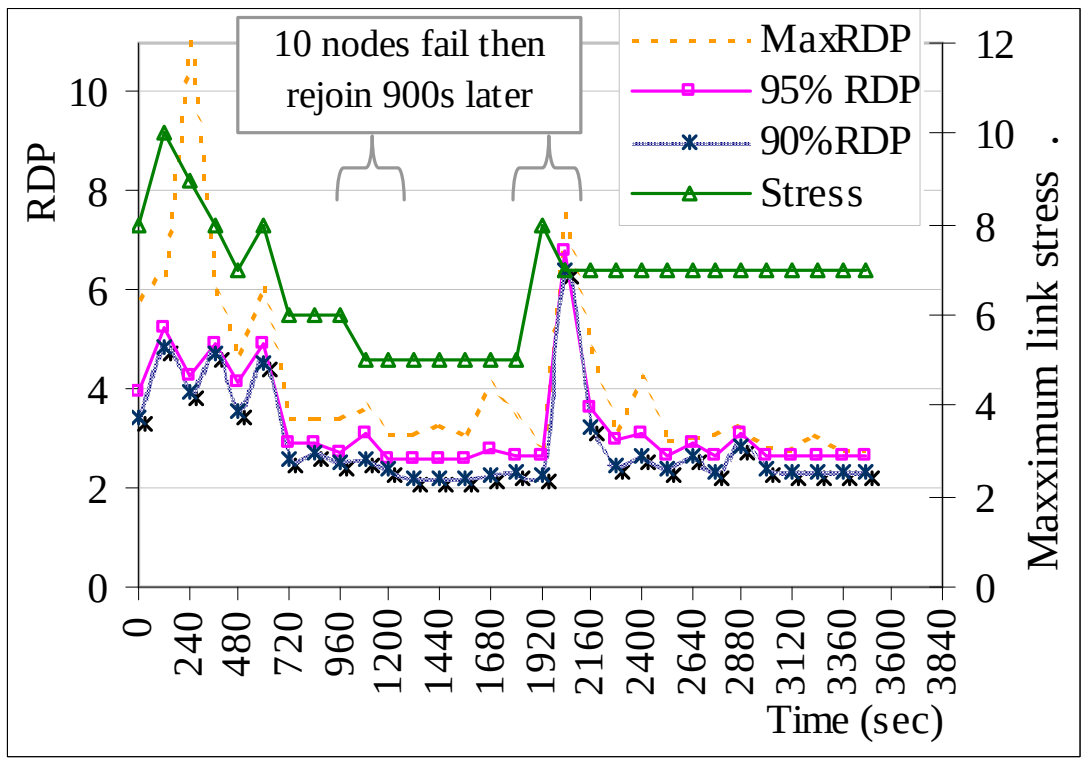

**Figure 12**: Adaptability. RDP and maximum link stress evolution for a multicast tree rooted at one random node. 900s after the 64-node overlay is started, 10 nodes fail during a 120s interval. The 10 nodes then rejoin 900s later, again during a 120s interval.

### 3.) *Agility: Repair Time*

Using the same physical topology and a 64-node overlay, we explore UMM's ability to operate in a dynamic environment. We consider a case when 10 nodes fail then rejoin the overlay 900s later. Both failure and join events are distributed over a 120s interval: seconds 1000 to 1120 for failures and 2020 to 2140 for joins on the timescale in Figure 12 presents the evolution of RDP and maximum link stress. 900s after bootstrap, RDP and stress are already close to their stable values. Since node failures are distributed over a relatively large interval, the base overlay recovers from these failures quickly and the RDP increase is not significant. Maximum link stress actually decreases as fewer participating nodes lead to fewer tunnels mapped to physical network links. The impact of nodes joining in is larger as new nodes take time to discover to which existing nodes to create overlay tunnels (Figure 12).

## VII OTHER RELATED WORK

In addition to the overlay-based multicast solutions presented in Section III, we acknowledge related work in two other contexts: flood-and-prune techniques and high-bandwidth data-distribution.

### *A. Flood-and-Prune Techniques*

Flood-and-prune techniques have been used at layers 2 and 3 in the network stack with various degrees of success and reflecting specific deployment assumptions.

At the *data link layer* (OSI layer 2), bridges employ a "spanning tree protocol" [42] to extract a shared spanning tree from a non loop-free local network. UMM's solution is different in two respects: first, the trees constructed are source-specific while the algorithm above builds a shared tree. This way UMM offers better use of resources (e.g., transport capacity, delay) available at the lower layer at the cost of maintaining additional state at each node. Efficient use of these resources is of lower concern in LAN deployment scenarios. Second, UMM builds trees by passively listening to actual traffic, without using per-tree control messages. UMM's solution, based on caches of recently received message identifiers, is enabled by larger memory available at end-hosts and application-level frames, which enable a coarser granularity than IP packets.

At the *network layer*, two commonly used multicast routing protocols, DVMRP (the Distance Vector Multicast Routing Protocol [23]) and DM-PIM (Dense-Mode Protocol Independent Multicast [43]), use a flood-and-prune technique on top of already selected multicast dissemination trees. The main difference when compared with UMM lies in deployment assumptions: DVMRP and DM-PIM are deployed across the Internet to extract single-source group-specific dissemination trees. This way, all Internet routers participate in all the tree extraction mechanisms (which implies flooding and holding state for pruning) for all groups, even when routers do not serve any end-hosts interested in these groups. UMM, in contrast, builds per group overlays and state is maintained at only participating end-hosts: thus, as for all application-layer multicast solutions, state is moved to the edges of the network and hosts pay a direct cost only when participating.

*B. High-Bandwidth Data Distribution*

A number of applications target high-bandwidth data distribution. Both Bullet [22] and SplitStream [44] assume a single source scenario.

Bullet establishes multiple trees rooted at the same source and disseminate different chunks of data. By using multiple trees, Bullet reduces the need to perform expensive bandwidth probing for tree optimization. Bullet's solution works best for single source distribution of large-files: it assumes that the application is insensitive to additional transfer delays, that delivery order of various data blocks is not important, and that the additional overhead incurred by coordinating multi-path data delivery is amortized over a large file size. UMM is designed for multiple source applications and assumes perishable data (e.g., conferencing, media streaming) in context where additional delays and message re-ordering are important concerns.

The motivating assumption for SplitStream [44] is that, in a cooperative scenario, nodes should have similar loads and, as a result, a distribution tree is not the appropriate solution since interior tree nodes bear the distribution load. Thus, SplitStream employs a solution similar to Bullet's: it uses multiple distribution trees, this time based on a structured overlay. UMM assumptions are different: we believe that since participating hosts are inherently heterogeneous a cooperative solution should assign load proportional to node capabilities. It is possible to add, on top of this, a system that rewards the high volume nodes.

# VIII Conclusions

Multicast communication primitives are building blocks for a large variety of distributed applications. However, creating and maintaining the distributed structures that support these primitives is challenging, particularly when network and node characteristics are transient.

UMM, the distributed multi-source multicast system we have proposed, is based on a simple approach to extracting source-specific multicast trees from an unstructured overlay. This solution offers two important properties: first, it decouples the overlay construction and maintenance mechanism from the tree-extraction mechanism, allowing for separate component optimization. Second, it relies on soft-state and passive data collection to adapt to the dynamics of the physical network, resulting in low protocol complexity and low overheads. Experimental and analytical evaluations demonstrate low communication overhead, efficient network usage compared to alternative solutions, and ability to adapt quickly to network changes and recover from node failures.

More importantly, we show that a low-complexity design can lead to a self-organizing, scalable, and adaptive overlay with performance generally better than that offered by more sophisticated solutions and, additionally, with low overheads compared to idealized solutions based on global resource views. Reduced complexity is a highly desirable property of distributed, large-scale systems. When two mechanisms offer similar efficiency for comparable costs, the discriminating factor is complexity. Additionally, in production environments an increasing premium is placed on deployable, manageable, in other words *simple* systems.

## X APPENDIX

The appendix presents pseudocode for the two UMM layers: base overlay management and dissemination tree extraction and maintenance.

### A. *Base overlay management.*

This section presents the pseudocode for the main thread in charge with the base overlay management

```
while (true) do:
    connCandidateSet = randomCandidates (MAX_CANDIDATES)
    foreach conn in connCandidateSet do:
        // probe the network path for this possible connection
        // and use exponentially weighted averages add with older data
        evaluatePotentialConnection (conn)

    // order the connection set in decreasing order of their potential
    sort(connCandidateSet)

    foreach conn in connCandidateSet  do:
        // try to replace the worst existing short connection
        // uninitialized local tunnels have infinite delay
        if isShort(conn) then:
            if (conn.delay < worstCurrShortTunnel.delay –
                                    DELAY_THRESHOLD)
    then:
                if (createNewConnection (conn)) then:
                    adjustConnectionsNumber()
                    delete (worstCurrShortTunnel)
                    break;

        // try to replace the worst existing long connection
        // uninitialized local tunnels have 0 bandwidth
        if isLong(conn) {
            if (conn.bw < worstCurrLongConn.bw *
                                BANDWIDTH_THRESHOLD)
then:
                if (createNewConnection (conn)) then:
                    adjustConnectionsNumber();
                    delete (worstCurrShortTunnel)
                    break;
    wait (OPTIMIZER_THREAD_DELAY)
end while
```

### B. *Dissemination tree extraction and maintenance*

At this layer each node uses three soft-state state data stores. These data stores are soft-state in the sense that data items are inserted with associated timeouts. When a timeout expires, the data store simply flushes the associated data item.

- A data store for IDs of received messages and the tunnel they arrived: idStore (id, conn, timeout)
- A data store for tunnels that have been filtered out by remote nodes: tunnelFiltered (source, conn, timeout)
- A data store for tunnels that have been filtered out by local nodes: tunnelFilteredLocal (source, conn, timeout)

The pseudocode blow describe UMM's reaction to various events:

```
event dataMessage(m, conn):
    oldConn = idStore.put(m.ID, conn, TIMEOUT_IDSTORE)
    // if this was a new ID put into the idStore
    if (oldConn == NULL) then:
        // then deliver to application and propagate the msg.
        callCallback (deliverMessageCallback, m)
        if (m.ttl – – <= 0) then:
            return
        foreach c in localNodeConnections do:
            if (not c in tunnelsFiltered) and (c != conn) then:
                send (m, c)
    else { // this was a duplicate so:
    // ask the node at the other of the connection not to send
    // messages from the same source again
    // (if the connection is active)
    if oldConn.active() and (oldConn not in tunnelFilteredLocal) then:
        send (new dropRouteMessage (m.scr), conn)
        tunnelFilteredLocal.put (m.scr, conn,
    TIMOUT_TUNNEL)

event dropRouteMessage (m, conn):
    tunnelsFiltered.put (m.scr, conn, TIMOUT_TUNNEL)

event failedConnection (conn):
    // when a tunnel failure is detected then spread a
    resetRouteMessage
    // with a small time to live
    m = new resetRouteMessage (RESET_TTL)
    foreach c in localNodeConnections do:
        send (m, c)
        tunnelFilteredLocal.reset(c)
    tunnelFiltered.reset()

event resetRouteMessage (m, conn):
    // only for new messages
    if (NULL == idStore.put(m.ID, conn,
TIMEOUT_IDSTORE)) then:
        tunnelFiltered.reset(conn)
        if (m.ttl – – <= 0) then:
            return
        foreach c in localNodeConnections do:
            send (m, c)
            tunnelFilteredLocal.reset(c)
```